\documentclass{appolb}
\usepackage{epsfig}

\begin{document}
\title{The $\sigma$ meson in a nuclear medium through two
pion photoproduction
}
\author{{L.~Roca, E.~Oset and M.~J.~Vicente Vacas}
\address{Depto. de F\'{\i}sica Te\'orica and IFIC. 
Centro Mixto Universidad de Valencia-CSIC}%
}
\maketitle
\begin{abstract}
We present theoretical results for $(\gamma,\pi^0\pi^0)$ and
$(\gamma,\pi^{\pm}\pi^0)$
 production
 on nucleons and nuclei
in the kinematical region where the scalar isoscalar $\pi\pi$ amplitude is 
influenced by the
$\sigma$ pole. The final state interaction of the pions modified by the nuclear 
medium produces a spectacular shift
of strength of the $\pi^0\pi^0$ invariant mass distribution induced by
 the moving of the $\sigma$ pole to lower masses and  widths as the nuclear 
 density increases.
\end{abstract}

\vspace{+0.5cm}
  
In the last years there has been an intense theoretical and
experimental debate about the nature of the $\sigma$ meson,
especially the modification of its properties in nuclear
matter. Both the interpretation as a $q\overline{q}$ state or
as a $\pi\pi$ resonance, seem to agree in a decrease of the
$\sigma$ mass as the nuclear density increases. For instance,
in the theoretical work of \cite{Chiang:1998di},
the $\pi\pi$ amplitude in the $\sigma$ channel 
showed a shift of strength towards low $\pi\pi$ invariant
masses at finite nuclear density.
 In the present work,
our purpose is to find the possible experimental
signature of this moving of the $\pi\pi$ pole in the
 $(\gamma,\pi^0\pi^0)$ reaction in nuclei.
This reaction is much better suited  than the
$(\pi,\pi\pi)$ one to investigate the
modification of the $\pi\pi$ in nuclear matter
because the photons are not distorted
by the nucleus and one can test higher densities.

For the model of the elementary $(\gamma, \pi \pi)$ reaction we follow 
 \cite{Nacher:2000eq} which considers the coupling of the photons to mesons, 
 nucleons, and the resonances
$\Delta(1232)$, $N^*(1440)$, $N^*(1520)$ and $\Delta(1700)$.
This model relies upon tree level
diagrams. Final state interaction of the $\pi N$ system is accounted for
by means of the explicit use of resonances with their widths. However,
since we do not include explicitly the $\sigma$ resonance, the final state
interaction of the two pions has to be implemented to generate it.

The $\gamma N \to N \pi^0 \pi^0$ amplitude can be decomposed in
a part which has in the final state the combination of pions
in isospin I=0 and another part where the pions are in I=2.
  
The renormalization of the 
$I=0$ $(\gamma,\pi \pi)$ amplitude is done by factorizing the on shell 
tree level $\pi N \to \pi \pi N $ and $\pi \pi \to \pi \pi$ amplitudes in the 
loop functions. 

\begin{equation}
T_{(\gamma,\pi^0\pi^0)}^{I_{\pi\pi}=0}\to T_{(\gamma,\pi^0\pi^0)}^{I_{\pi\pi}=0}
\left(1+G_{\pi\pi}t_{\pi\pi}^{I=0}(M_I)\right)
\label{eq:GT1}
\end{equation}
where $G_{\pi\pi}$ is the loop function of the two pion propagators, 
which appears in the Bethe Salpeter equation, and $t_{\pi\pi}^{I=0}$ is the
$\pi\pi$ scattering matrix in isospin I=0, taken from \cite{Chiang:1998di}.
In the model for $( \gamma,2\pi )$ of 
\cite{Nacher:2000eq} there are indeed contact terms as implied before, as well as
other terms involving intermediate nucleon states or resonances. In this
latter case the loop function involves three propagators but 
the intermediate baryon is far off shell an the factorization of
Eq.~(\ref{eq:GT1}) still holds. There is, however, an exception in
the $\Delta$ Kroll Ruderman term, since as we increase the photon energy we get
closer to the $\Delta$ pole. Fort this reason this term has
been dealt separately making the
explicit calculation of the loop with one $\Delta$ and two pion propagators.

The cross section for the process in nuclei is calculated using many body techniques.
The distortion of the final pions in their way out through the nucleus
via absorption and
quasielastic collisions is of crucial
importance and is taken into account by using an eikonal approximation.
The probability of
absorption and quasielastic collisions is calculated using a
well tested model for the pion selfenergy
in nuclear matter for low energy pions. With this approximation
the pions which undergo absorptions are removed
from the flux but we do not remove those which undergo 
quasielastic collisions since they do not change in average
the shape or the strength of the $\pi\pi$
 invariant mass distribution.

In the figure we can see the results for the two pion invariant mass distributions 
in the $(\gamma,\pi^0\pi^0)$ and $(\gamma,\pi^{\pm}\pi^0)$ reactions
 on $^{12}C$ and $^{208}Pb$.
 The difference between the 
solid and dashed curves is the use of the in
medium $\pi \pi$ scattering and $G$ function instead of the free ones, which
we take from \cite{Chiang:1998di}.  As one can see in the figure, there 
is an appreciable shift of strength to the low
invariant mass region in the $(\gamma,\pi^0\pi^0)$
due to the in medium  $\pi \pi$  interaction.   This
shift is remarkably similar to the one found in the  measurements of
\cite{Messchendorp:2002au}.
This shift is not seen in the $(\gamma,\pi^{\pm}\pi^0)$ channel because the
$\pi^{\pm}\pi^0$ are not allowed to be in isospin I=0.

These results show a clear signature of the modified
$\pi\pi$ interaction in the nuclear medium that can
be related to the modification of the $\sigma$
meson in the medium since, in the model
we use, the $\sigma$ meson is dynamically 
generated by the multiple scattering of the pions
through the underlying chiral dynamics.

\newpage

\begin{figure}
\includegraphics[angle=0,width=10cm]{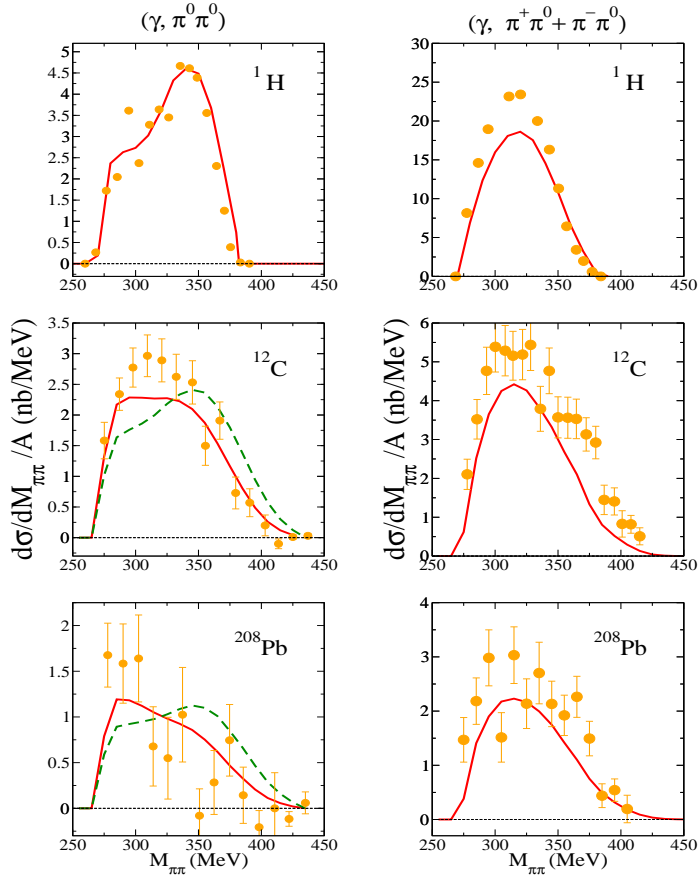}
\caption{\small{Two pion invariant mass distribution for $\pi^0\pi^0$ and $\pi^{\pm}\pi^0$
 photoproduction 
in $^{12}C$ and $^{208}Pb$.
Continuous lines: using the in medium final $\pi\pi$ interaction. 
Dashed lines: using the final $\pi\pi$ interaction at zero
density.
Exp. data from \cite{Messchendorp:2002au}}}
\end{figure}

\end{document}